# Nanoscale Strain Engineering of Giant Pseudo-Magnetic Fields, Valley Polarization and Topological Channels in Graphene

C.-C. Hsu,[1]† M. L. Teague,[1]† J.-Q. Wang,[1]† N.-C. Yeh[1,2]*

[1]Department of Physics, California Institute of Technology, Pasadena, CA 91125, USA
[2]Kavli Nanoscience Institute, California Institute of Technology, Pasadena, CA 91125, USA

*Corresponding author: Nai-Chang Yeh (ncyeh@caltech.edu)
†These authors contributed equally to this work

**Abstract**

The existence of nontrivial Berry phases associated with two inequivalent valleys in graphene provides interesting opportunities for investigating the valley-projected topological states. Examples of such studies include observation of anomalous quantum Hall effect in monolayer graphene, demonstration of topological zero modes in "molecular graphene" assembled by scanning tunneling microscopy, and detection of topological valley transport either in graphene superlattices or at bilayer graphene domain walls. However, all aforementioned experiments involved non-scalable approaches of either mechanically exfoliated flakes or atom-by-atom constructions. Here we report an approach to manipulating the topological states in monolayer graphene *via* nanoscale strain engineering at room temperature. By placing strain-free monolayer graphene on architected nanostructures to induce global inversion symmetry breaking, we demonstrate the development of giant pseudo-magnetic fields (up to ~ 800 Tesla), valley polarization, and periodic one-dimensional topological channels for protected propagation of chiral modes in strained graphene, thus paving a pathway towards scalable graphene-based valleytronics.

**One sentence summary:** Nanoscale strain engineering of monolayer graphene is shown to achieve giant pseudo-magnetic fields and valley polarization.

**MAIN TEXT**

## Introduction

It has been well recognized that the Berry curvature of electronic wave functions can have a profound effect on the physical properties of materials.[1-3] For instance, non-trivial Berry curvatures in the event of either broken time-reversal symmetry or broken inversion symmetry are known to be responsible for various (quantum, anomalous, spin and valley) Hall effects.[3-6] In the case of graphene with gapless Dirac cones at the two inequivalent valleys K and K′, the spinor-type wavefunctions of the Dirac fermions result in non-trivial Berry phases of $\pi$ and $\pi-$.[7-9] For perfect and flat monolayer graphene in the absence of an external magnetic field, the Berry flux from K and K′ exactly cancels each other so that the Hall conductance vanishes under both time and inversion symmetries.[7-9] On the other hand, inversion symmetry can be broken by either atomically aligning monolayer graphene on top of $h$-BN,[10] or artificially building in the broken inversion symmetry in strained "molecular graphene" assembled by scanning tunneling microscopy.[11] The formal leads to the realization of the valley Hall effect (VHE)[10] and the latter manifests Landau quantization and site-dependent topological zero modes in the tunneling conductance spectra.[11] Additionally, one-dimensional valley-polarized conducting channels associated with the protected chiral edge states of quantum valley Hall insulators have been demonstrated at the domain walls between AB- and BA-stacked bilayer graphene.[12] These interesting results underscore the rich opportunities provided by graphene-based systems for the studies of valley-projected topological states.

In contrast, while monolayer transition-metal dichalcogenides (TMDCs) in the 2H-phase are two-dimensional (2D) semiconducting crystals with an in-plane structure similar to graphene and exhibit strong spin-valley coupling and interesting optical properties,[13-16] no discernible out-of-the-plane pseudo-magnetic fields can be induced by strain due to negligible Berry curvatures around the 60 gapped bands at the K and K′ points.[17] Thus, the primary strain-induced effect on 2D-TMDCs is only associated with the modification to the semiconducting energy gaps, leading to brighter photoluminescence in stronger strained areas due to the resulting smaller energy gaps.[17,18]

Earlier experimental investigations of strained graphene generally involve approaches of either stacking mechanically exfoliated, microscale flakes of graphene on $h$-BN,[10] or assembling atom by atom using a scanning tunneling microscope.[11] Clearly neither method is scalable for realistic device applications. Recently, efforts have been made to pre-design the substrate to induce controlled strain on monolayer graphene.[19-24] However, most of these strained graphene samples have only been characterized by imaging with scanning electron microscopy (SEM) and/or atomic orce microscopy (AFM) and by Raman spectroscopy without explicit investigation of the strain-induced pseudo-magnetic field.[19-23] In the only case of direct measurements of the pseudo-magnetic fields by means of scanning tunneling microscopy/spectroscopy (STM/STS), studies were solely carried out in nearly flat areas with much smoother topography so that the resulting pseudo-magnetic fields were very small (~ 7 Tesla) and were taken at low temperatures (~ 4.6 K).[24] Here we report a scalable experimental approach that successfully induces giant pseudo-magnetic fields (up to ~ 800 Tesla) and achieves manipulation of the topological states and in monolayer graphene by means of nanoscale strain engineering *at room temperature*. Our methodology involves placing a nearly strain-free, large-area (~ 1 cm$^2$) monolayer graphene sheet on top of properly architected nanostructures to induce global inversion symmetry breaking. We demonstrate the development of strain-induced giant pseudo-magnetic fields and global valley polarization by direct STM/STS studies. The experimental investigations are further corroborated by simulations using the Molecular Dynamics (MD) method elaborated in Supplementary Material and Figs. S1-S3. We also verify the feasibility of periodic one-dimensional topological channels for protected propagation of chiral modes in graphene based on our empirically demonstrated periodic strain patterns and MD simulations. This methodology is shown to be scalable and controllable, which paves a new 86 pathway towards realizing realistic graphene-based valleytronic applications.

## Results

*Nanoscale strain engineering of graphene*

   Our experimental approach is based on the notion that the electronic properties of graphene exhibits significant dependence on the nanoscale structural distortion and the resulting strain[7,25–31]. In general, structural distortion-induced strain in the graphene lattice gives rise to two primary effects on the Dirac fermions.[13,15] One is an effective scalar potential $\Phi$, and the other is an effective gauge potential $\mathcal{A}$ related to the Berry connection in the reciprocal space.[25,27] These strain-induced effective scalar and gauge potentials in graphene are the consequences of the changes in the distance or angle between the $p_z$ orbitals due to structural distortions, which modify the hopping energies between Dirac electrons at different lattice sites, thereby giving rise to the addition of an effective gauge potential $\mathcal{A}$ and a scalar potential $\Phi$ to the original Dirac Hamiltonian of ideal monolayer graphene. The excess scalar potential can cause scattering of Dirac fermions and changes in the local charge densities, whereas the excess gauge potential can result in a pseudo-magnetic field that is related to the Berry curvatures in the reciprocal space and couples with the valley (pseudo-spin) degrees of freedom. Thus, proper nanoscale engineering of the strain in graphene can provide unique means to manipulate the valley degrees of freedom.[29] Moreover, by combining large-scale strain-free graphene and modern nanofabrication technology, it becomes feasible to devise scalable structures to achieve desirable control of the valley-associated topological states.

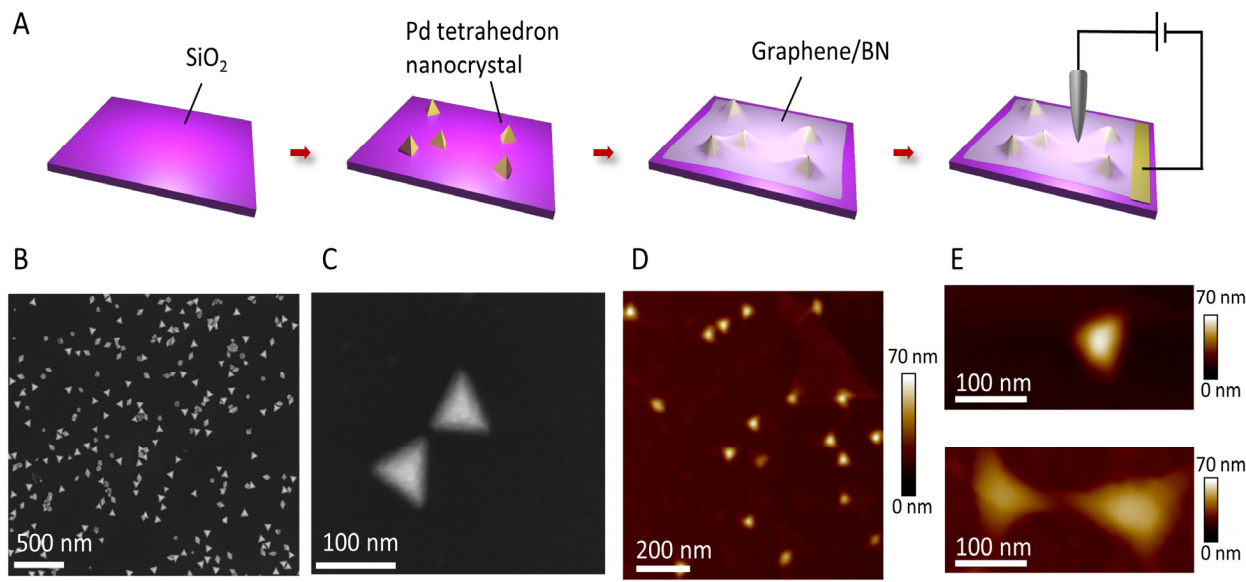

**Fig. 1. Nanoscale strain engineering of graphene:** (**A**) Schematic illustrations showing the steps taken to induce strain on graphene by Pd tetrahedron nanocrystals (NCs). (**B**) An SEM image of randomly distributed Pd tetrahedron NCs distributed on a Si substrate over a (3 × 3) μm$^2$ area. (**C**) A zoom-in SEM image of two Pd tetrahedron NCs. (**D**) Exemplifying AFM image of graphene/$h$-BN/Pd tetrahedron NCs. (**E**) <u>Top panel</u>: AFM image of graphene/BN on a single Pd tetrahedron NC, showing excellent conformation of graphene/BN to the single Pd tetrahedron NC. <u>Bottom panel</u>: AFM image of graphene/BN on two closely spaced Pd tetrahedron NCs, showing the formation of a graphene "wrinkle" between the two Pd tetrahedron NCs.

To achieve large-area (~ 1 cm$^2$) nearly strain-free graphene, we employed the plasma enhanced chemical vapor deposition (PECVD) technique, as detailed previously.[32] To induce controlled nanoscale strain on initially strain-free graphene, we carried out the procedures schematically illustrated in Fig. 1A. Specifically, two different approaches were taken to induce substantial strain. The first approach involved fabricating Pd nanocrystals (NCs) that were in the form of tetrahedrons with a typical base of ~ 30 nm in length, using a wet chemical method first developed by Y. Zhang *et al.*[33] and further elaborated in Materials and Methods. The second approach involved fabricating periodic arrays of nanostructures on silicon substrates using either electron beam (e-beam) lithography or focused ion beam (FIB) lithography, as detailed in Materials and Methods and further illustrated in Fig. S4. The NCs were dispersed on a silicon substrate, as exemplified by the SEM images over an area of (3 × 3) μm$^2$ shown in Fig. 1B and for a zoom-in view of two NCs shown in Fig. 1C. Next, we transferred a monolayer of $h$-BN and then a monolayer of our PECVD-grown monolayer graphene on top of the NCs by means of a polymer-free technique,[34] where the h-BN was either mechanically exfoliated from an h-BN single crystal or by the CVD growth method 143 developed by W.-H. Lin et al.[35] Both the h-BN and graphene monolayers were shown to conform well to separated NCs, as illustrated in Fig. 1D and the upper panel in Fig. 1E. Interestingly, however, for closely spaced NCs, a wrinkle in graphene appeared, as exemplified in the lower panel of Fig. 1E. Here we note that the purpose of inserting an h-BN monolayer between graphene and silicon nanostructures is to minimize the effects of phonon coupling between Si and graphene[28] and of charge impurities in the Si substrate[28] on graphene.

The structural distortion-induced strain on graphene can be evaluated using the strain tensors. For a spatially varying displacement field $\mathbf{u} \equiv u_x \hat{x} + u_y \hat{y} + h\hat{z}$, the strain tensor components $u_{xx}$, $u_{xy}$ and $u_{yy}$ are given by[25,27]

$$u_{xx} = \frac{\partial u_x}{\partial x} + \frac{1}{2}\left(\frac{\partial h}{\partial x}\right)^2, \quad u_{xy} = \frac{1}{2}\left(\frac{\partial u_x}{\partial y} + \frac{\partial u_y}{\partial x}\right) + \frac{1}{2}\left(\frac{\partial h}{\partial x}\frac{\partial h}{\partial y}\right), \quad u_{yy} = \frac{\partial u_y}{\partial y} + \frac{1}{2}\left(\frac{\partial h}{\partial y}\right)^2. \quad (1)$$

If the x-axis is chosen along the zigzag direction, the two-dimensional strain-induced gauge potential in the real space $\mathcal{A} = A_x \hat{x} + A_y \hat{y}$ can be expressed in terms of the tensor components by the following relations (in the first order approximation):[7]

$$A_x = \pm \frac{\beta}{a_0}(u_{xx} - u_{yy}), \quad A_y = \mp 2\frac{\beta}{a_0}u_{xy}, \quad (2)$$

where $a_0 \approx 0.142$ nm is the nearest carbon-carbon distance for equilibrium graphene, $\beta$ is a constant ranging from 2 to 3 in units of the flux quantum, and the upper and lower signs are associated with the K and K′ valleys, respectively.[7] From Eq. (1), we note that in the event of strong z-axis corrugation, the strain components resulting from the height variations becomes dominant over the in-plane strain components, as exemplified in Figs. S1-S3. Using Eqs. (1) and (2), the spatial distribution of the pseudo-magnetic field can be obtained according to $B_S(\mathbf{r}) = \times \nabla \mathcal{A}(\mathbf{r})$, with opposite signs associated with the K and K′ valleys so that the global time reversal symmetry is preserved and total flux integrated over the entire sample is zero. Hence, from given three-dimensional atomic structural distortions of graphene, which may be determined by either scanning tunneling microscopy (STM) or high-resolution AFM, the spatial variations of $B_S(\mathbf{r})$ can be derived.

Alternatively, the magnitude of pseudo-magnetic field, $|B_S(\mathbf{r})|$, can be independently verified by spatially resolved scanning tunneling spectroscopy (STS), where the pseudo-magnetic field-induced quantized Landau levels $E_n$ (with $n$ being integers) in the tunneling conductance ($dI/dV$) vs. biased voltage ($V = E/e$, with $E$ being quasiparticle energy) spectrum at a given position $\mathbf{r}$ satisfy the following relation:

$$E_n = \mathrm{sgn}(n)\sqrt{2ev_F^2\hbar|nB_S|}, \quad \Rightarrow \quad |B_S| = \left[(E_{n+1})^2 - (E_n)^2\right]/(2ev_F^2\hbar). \quad (3)$$

Using Eq. (3), the magnitude of the pseudo-magnetic field at a given position can be determined rigorously by the energy spacing of different Landau levels of varying indices $n$ in the local tunneling spectrum. The consistency of such *spectroscopic* studies with the value obtained from the strain tensors can be verified by comparing with the atomically resolved *topographic* studies.

## Topographic and spectroscopic evidences for the formation of giant pseudo-magnetic fields

In Fig. 2 we illustrate the comparison of the strain-induced pseudo-magnetic fields for the K-valley from both topographic and spectroscopic studies at room temperature. The main panel of Fig. 2A and Fig. 2B are respectively zoom-out AFM and STM topographic images over a (100 × 100) nm$^2$ 188 area that cover the full view of monolayer graphene/$h$-BN over an isolated Pd-tetrahedron. In the inset of Fig. 2A, a zoom-in atomically resolved STM topography of graphene over a (3 × 3) nm$^2$ area near the tip of the tetrahedron reveals strong structural distortion in graphene with significant height displacements. Assuming the validity of first-order strain-induced perturbation to the Dirac Hamiltonian and using the Molecular Dynamics (MD) method as detailed in Supplementary Material, we obtain the resulting pseudo-magnetic field distributions in Fig. 2C for the topography shown in Fig. 2B. Additionally, maps of the corresponding strain tensors are provided in Figs. S5A-S5C. Given the significant structural distortions in graphene, we note the resulting large magnitudes of the pseudo-magnetic field, up to ~ 800 Tesla in maximum values if computed from the topographic information.

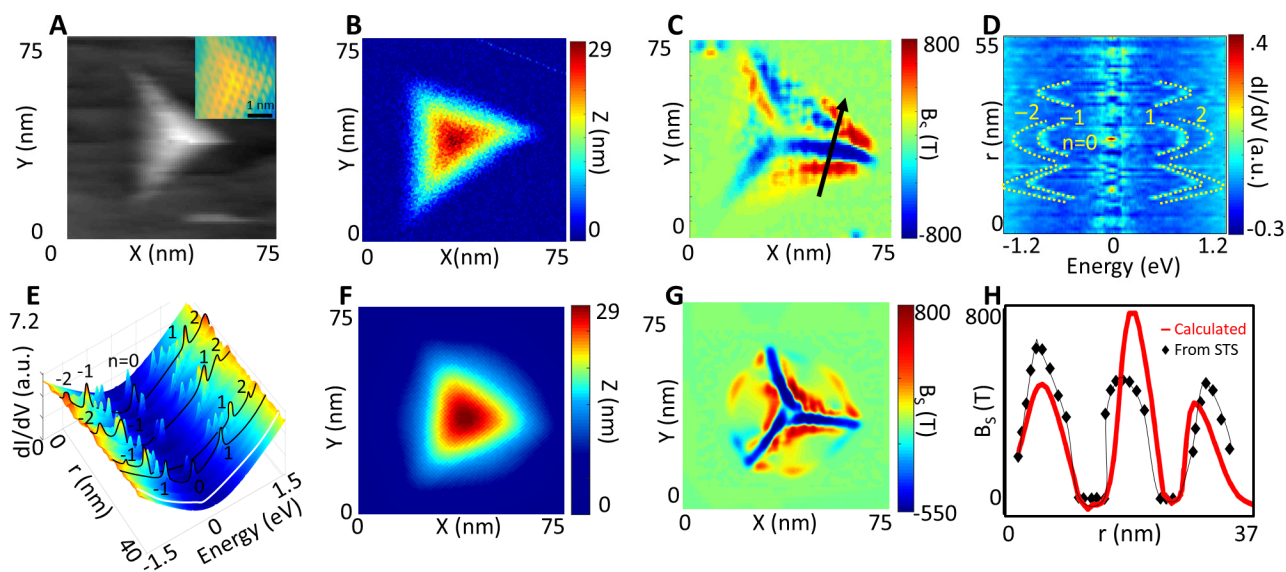

**Fig. 2. Topographic and spectroscopic studies of strain-induced effects on graphene at room temperature due to one Pd tetrahedron NC:** (**A**) Three-dimensional (3D) topographic images of the distorted graphene taken by AFM (Main Panel) and by STM (Inset, zoom-in image with atomic resolution). (**B**) 3D topographic image of the distorted graphene taken by STM. (**C**) The pseudo-magnetic field map calculated from the topography over the same area as shown in (B). (**D**) Tunneling conductance spectral difference relative to the Dirac spectrum of strain-free graphene is shown along the line-cut indicated by the black arrow in (C), revealing spatially varying strengths of strain-induced pseudo-magnetic fields as manifested by the variations in the Landau-level separation. (**E**) Representative spectra of tunneling conductance-vs.-energy of strained graphene along the black line-cut in (C), showing quantized conductance peaks in strained regions and the V-shape Dirac spectrum in strain-free regions as exemplified by the white curve located at $r \sim 36$ nm. (**F**) 3D topographic map of graphene/$h$-BN deformation on an ideal tetrahedron, as computed from MD simulations described in Supplementary Material. (**G**) Pseudo-magnetic field map computed from the topographic distortion in (F). (**H**) Comparison of the absolute values of pseudo-magnetic fields $|B_s(r)|$ derived from topographic studies (red line) and from the Landau level separations in STS (black diamonds), showing overall satisfactory agreement. Here $r$ denotes the distance measured from the lower-left end to the upper-left end of the black arrow shown in (C).

Concurrent spectroscopic studies of the strained graphene over the isolated tetrahedron also revealed spatially varying tunneling spectra, as exemplified in Fig. 2D for a collection of high-resolution tunneling conductance vs. bias voltage spectra along the black line indicated in Fig. 2C. Here the horizontal axis in Fig. 2D corresponds to the bias voltage, the vertical axis corresponds to the spatial dimension along the black line (from lower left to upper right) in Fig. 2C, and the colors represent the tunneling conductance difference from the unstrained graphene. The three-dimensional representation of the tunneling spectra taken along the same line-cut are shown in Fig. 2E. Specifically, a typical V-shaped tunneling spectrum for ideal graphene is clearly shown in the strain-free region, as exemplified by the white curve in Fig. 2E, whereas increasing larger energy separations for consecutive peak features are found for the tunneling spectra taken at increasingly strained regions, showing a consistent increase in the Landau level energy separations with the increasing magnitude of strain found in the topographic studies.

To further verify the consistency between the magnitude of pseudo-magnetic field determined from topography and from spectroscopy, we compare in Fig. 2H the absolute values of pseudo-magnetic fields $|B_S(r)|$ derived from topographic studies (Fig. 2C) and those from the Landau level separations (Fig. 2D) using Eq. (3), and find overall reasonable agreement. Here $r$ denotes the distance measured from the lower-left end to the upper-left end of the black arrow in Fig. 2C. Additionally, we carried out MD simulations for the topography and pseudo-magnetic field map of monolayer graphene/$h$-BN strained by a perfect tetrahedron with a base dimension of 30 nm, as shown in Figs. 2F and 2G, respectively. These MD simulations are largely consistent with the experimental results shown in Figs. 2B and 2C, although it is difficult to achieve detailed agreement due to unknown microscopic interaction parameters between the monolayer graphene/$h$-BN and the underlying nano-tetrahedron that are required to carry out the MD simulations.

To better manifest the characteristics of point spectra taken on areas of strained graphene, we show in Fig. S6A four point spectra taken on highly strained locations indicated as $\gamma, \beta, \alpha$ and $\delta$ on the pseudo-magnetic field map in Fig. S6B (which is the same pseudo-magnetic field map as Fig. 2C), where the corresponding pseudo-magnetic fields are $|B_S(r)| \sim 600$ Tesla and the resulting Landau levels $n = 0, \pm 1, \pm 2$ and $\pm 3$ are explicitly indicated. Additionally, a theoretical fitting curve for one of the point spectra with $|B_S(r)| = 592$ Tesla is shown in Fig. S6C, demonstrating that the superposition of Lorentzian Landau levels on top of a background Dirac spectrum achieves good agreement with the experimental data. We further note that for all point spectra taken in strained graphene areas, approximately a half of the spectra reveal a zero-bias conductance peak that corresponds to a Landau level with $n = 0$, whereas the other half of the spectra are without a zero-bias Landau level. This phenomenon is the result of two zero modes associated with spontaneous local time-reversal symmetry breaking, which will be investigated further in the Discussion section.

Next, we consider the strain on graphene induced by two closely spaced nano-tetrahedrons, as manifested by the topography in Figs. 3A and 3B and the corresponding pseudo-magnetic field map for the K-valley in Fig. 3C. Additionally, maps of the strain tensors associated with Fig. 3B are given in Figs. S7A-S7C. We found that the maximum magnitude for the pseudo-magnetic field computed from the structural distortion was $\sim 600$ Tesla (Fig. 3C), smaller than that found in the case of single tetrahedron (Fig. 2C). This is because comparable height displacements to those in Fig. 2B were spread over a larger lateral dimension in the case of two closely spaced tetrahedrons so that the magnitude of $(\partial h/\partial i)(\partial h/\partial j)$ becomes significantly reduced, where $i$ and $j$ denotes either $x$ or $y$ coordinate. Moreover, detailed comparisons of the spectroscopically determined pseudo-magnetic fields (as exemplified in Fig. 3D for the line-cut spectra along the white dashed line and in Figs. 3E and 3H for the line-cut spectra along the black dashed lines in Fig. 3C) with those determined topographically (Fig. 3C) were found to be in good agreement quantitatively.

In addition to verifying the consistency between the topographic and spectroscopic derivations of strain-induced pseudo-magnetic fields, the development of a topographic "wrinkle" between two nearby nanostructures is noteworthy. Moreover, the resulting pseudo-magnetic fields along the wrinkle direction appeared to form quasi-one dimensional "channels" of nearly uniform pseudo- magnetic fields, whereas those perpendicular to the wrinkle exhibited relatively rapid and continuous spatial variations with alternating signs. This formation of a topographic wrinkle in graphene between two nanostructures provides a hint for developing controlled and spatially extended strain to achieve global inversion symmetry breaking, which is the subject of our following exploration.

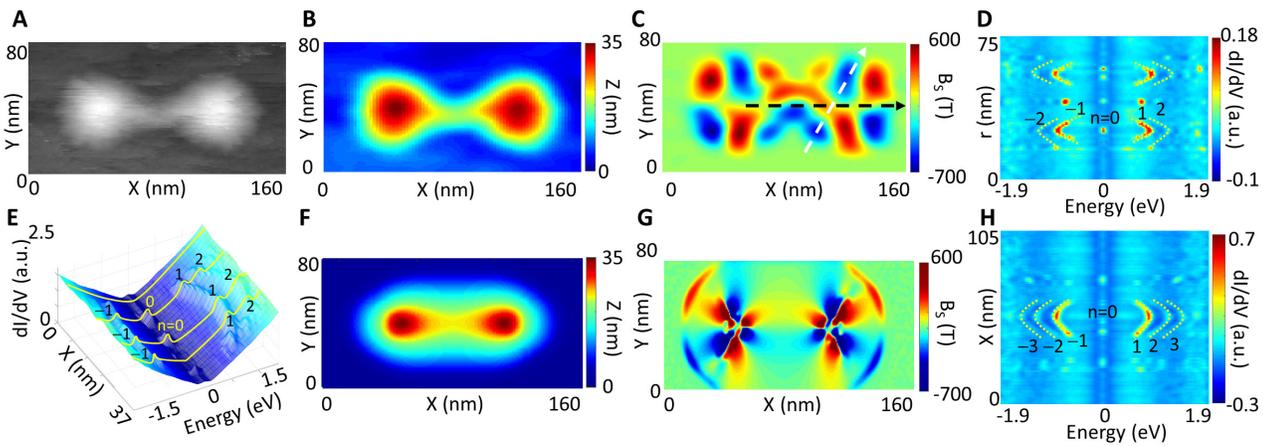

**Fig. 3. Topographic and spectroscopic studies of strain-induced effects on graphene due to two closely separated Pd tetrahedron NCs:** (**A**) Three-dimensional (3D) topographic image of the distorted graphene taken by AFM. (**B**) 3D topographic image of the distorted graphene taken by STM. (**C**) The pseudo-magnetic field map calculated from the topography over the same area as shown in (B). (**D**) The tunneling conductance spectral difference from the Dirac spectrum along the line-cut shown by the white dashed line in (C). (**E**) Spatially resolved tunneling spectra of strained graphene along the black dashed line in (C), showing strain-induced quantized conductance peaks. (**F**) 3D topographic map of graphene/$h$-BN on two ideal tetrahedrons computed from MD simulations. (**G**) Pseudo-magnetic field map computed from topographic distortion shown in (F). (**H**) The tunneling conductance spectral difference relative to the Dirac spectrum along the line-cut shown by the black dashed line in (C).

*Formation of periodic parallel graphene wrinkles for valley splitting and as topological channels*

Next, we employed nanofabrication technology to develop regular arrays of nano-cones on silicon with processes described in Materials and Methods and schematically illustrated in Fig. S4. Two types of periodic arrays were explored. One was a triangular lattice structure and the other was a rectangular lattice structure, as shown by the SEM images in the top panels of Figs. 4A and 4B, respectively. We found that the wrinkles induced on monolayer graphene by a triangular lattice had the tendency of forming along any of the three equivalent directions, as shown by the SEM image in the bottom panel of Fig. 4A. In contrast, wrinkles induced by the rectangular lattice were generally well aligned and parallel to each other, as exemplified by the SEM image in the bottom panel of Fig. 4B and the AFM images in the top panels of Figs. 4C and 4D. The corresponding pseudo-magnetic fields associated with the graphene distortions in the top panels of Figs. 4C and 4D are computed from the topography and shown in bottom panels of Figs. 4C and 4D.

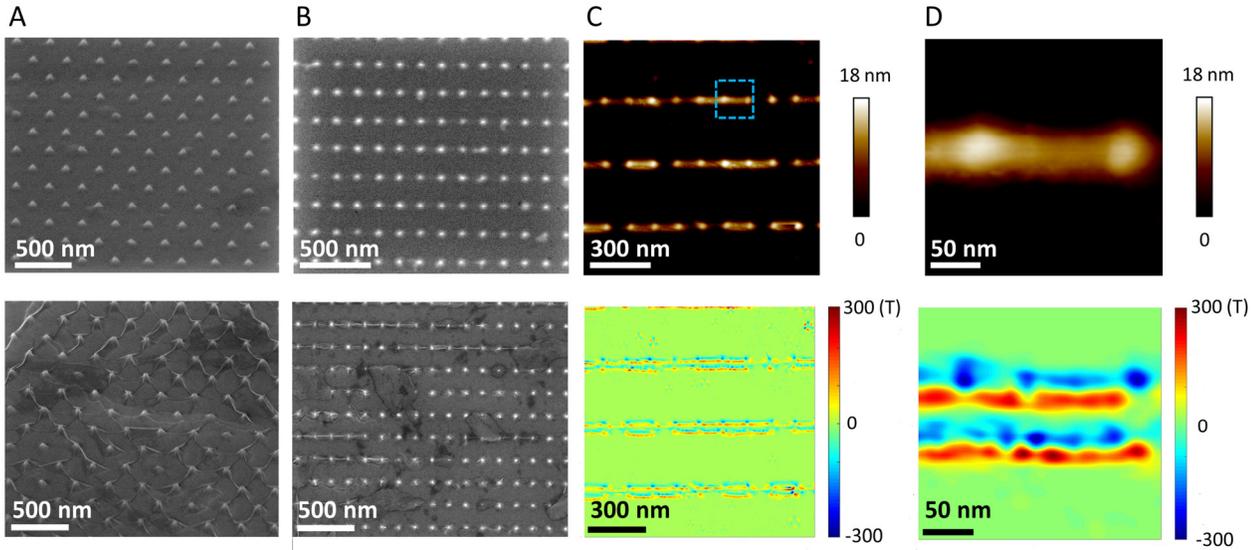

**Fig. 4. Extended strain effects induced by periodic arrays of nano-cones on graphene:** (**A**) Top panel: SEM image of triangular arrays of cone-shaped nanostructures fabricated on a SiO$_2$/Si substrate. Bottom panel: SEM image of monolayer-graphene/h-BN films on the triangular arrays shown in the top panel, showing graphene wrinkles appeared randomly along three equivalent directions. (**B**) Top panel: SEM image of rectangular arrays of cone-shaped nanostructures fabricated on a SiO$_2$/Si substrate. Bottom panel: SEM image of monolayer graphene/h-BN films on the rectangular arrays shown in the top panel, showing graphene wrinkles parallel to the axis of closer spaced nanostructures. (**C**) AFM image (top panel) of three parallel graphene wrinkles and the corresponding map of pseudo-magnetic fields derived from the strain tensors (bottom panel). (**D**) AFM image (top panel) of the graphene wrinkle enclosed by the blue dashed box in (C) and the corresponding map of pseudo-magnetic fields derived from the strain tensors (bottom panel).

It is worth noting that each extended graphene wrinkle results in four parallel, relatively uniform pseudo-magnetic fields along one direction and varying with alternating signs perpendicular to the channels, as illustrated in the bottom panel of Fig. 4D. Given that the pseudo-magnetic fields as observed by K and K' Dirac fermions are opposite in sign, the formation of parallel channels of pseudo-magnetic fields can effectively result in valley splitting and valley polarization. As illustrated by the theoretical simulations in the upper panels of Figs. 5A-5B and further detailed in Supplementary Material, for valley-degenerate Dirac fermions incident perpendicular to the parallel channels of pseudo-magnetic fields, K- and K'-valley fermions can become spatially separated and the lateral separation will increase with the increasing number of wrinkles they pass over, provided that the average separation ($d$) of consecutive wrinkles is less than the ballistic length ($l_B$) of Dirac fermions.

Specifically, the ballistic length $l_B$ is related to the conductance ($G$), mobility ($\mu$) and carrier density ($n_{2D}$) of Dirac fermions in monolayer graphene by the following relation:[7]

$$G = \frac{2e^2}{2\pi\hbar}(k_F l_B) = n_{2D} e\mu, \quad \Rightarrow \quad l_B = \left(\frac{2\pi\hbar}{2e}\right)\frac{n_{2D}\mu}{k_F} = \left(\frac{\hbar}{e}\right)\mu\sqrt{\pi n_{2D}}, \tag{4}$$

where $k_F = (n_{2D}\pi)^{1/2}$ is the Fermi momentum, $e$ is the electron charge, and $2\hbar\pi$ denotes the Plank constant. For typical values of $n_{2D} = 10^{10} \sim 10^{12}$ cm$^{-2}$ and $\mu \sim 10^5$ cm$^2$/V-s for our PECVD grown graphene, we find that $l_B = 120$ nm $\sim 1.2$ μm. Thus, by proper nanofabrication to design the $d$ value and by gating the PECVD-grown graphene for suitable $n_{2D}$ and $l_B$, the condition $d < l_B$ can be satisfied within realistic experimental parameters to achieve valley splitting and therefore valley polarized currents.

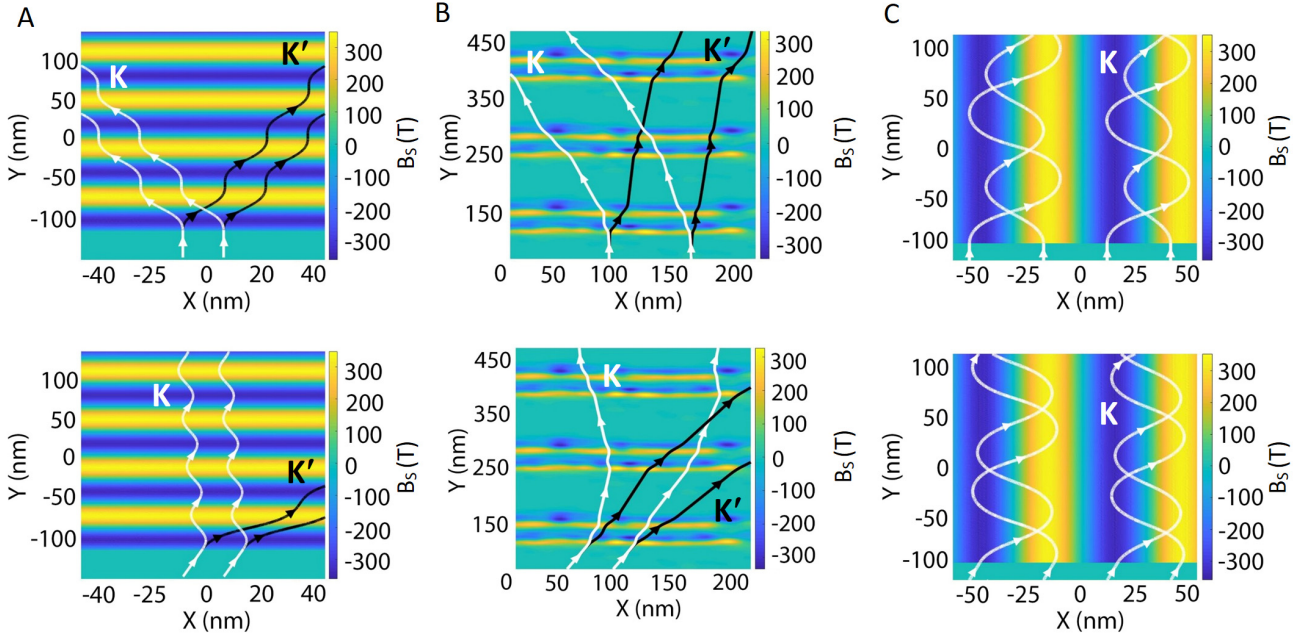

**Fig. 5. Parallel graphene wrinkles as topological channels for valley splitting and valley-polarized propagation:** (**A**) Simulations for parallel graphene wrinkles as a valley splitter, showing the trajectories of initially valley-degenerate (K + K′) fermions from strain-free regions becoming split when injected vertically into regions with strain-induced periodic channels of pseudo-magnetic fields. Top panel: Trajectories of K and K′ fermions for an incident angle perpendicular to the parallel channels ($\theta = 0°$). Bottom panel: Trajectories of K- and K′-valley fermions for an incident angle at $\theta = 15°$ relative to the normal vector of the parallel channels. (**B**) Top panel: Simulated trajectories of K and K′ fermions for an incident angle perpendicular to the realistic strain-induced parallel pseudo-magnetic fields ($\theta = 0°$) shown in Fig. 4C. Bottom panel: Simulated trajectories of K and K′ fermions for an incident angle at $\theta = 15°$ relative to the normal vector of the realistic strain-induced parallel pseudo-magnetic fields shown in Fig. 4C. (**C**) Simulations for parallel graphene wrinkles as a valley propagator, showing the collimation of valley-polarized fermions. Top panel: Trajectories of K-valley fermions incident at an angle parallel to the channels ($\theta = 90°$). Bottom panel: Trajectories of K-valley fermions incident at an angle $\theta = 75°$ relative to the normal vector of the parallel channels.

In addition to yielding valley splitting as discussed above, the parallel distributions of alternating signs of pseudo-magnetic fields can serve as *topological channels* for chiral Fermions. As shown in Fig. 5C, theoretical simulations for realistic arrays of nanostructures reveal that chiral Dirac fermions (*i.e.*, either K- or K′ fermions) can be preserved when propagate along the parallel channels of strain-induced pseudo-magnetic fields, as illustrated by the simulations shown in the top panel of Fig. 5C. Additionally, valley-polarized Dirac fermions can even be collimated along the topological channels if the incident angle deviates slightly from the channel direction, as exemplified in the bottom panel of Fig. 5C. Thus, parallel graphene wrinkles can serve as an effective conduit for protected propagation of valley-polarized Dirac fermions.

## Discussion

*Spontaneous local time-reversal symmetry breaking and the resulting two zero modes*

Although the strain-induced pseudo-magnetic fields do not break the global time-reversal symmetry, the gauge potentials $\mathcal{A}$ and $\mathcal{A}^*$ associated with the two valleys (*a.k.a.* two pseudo-spins) K and K′ in reciprocal space are opposite in sign and give rise to a peculiar zero mode.[36] This zero mode corresponds to a condensate where the Dirac fermions are delocalized over the entire sample, and yet they remain alternately localized and anti-localized for the pseudo-spin projection in the real space, yielding local spontaneous time-reversal symmetry breaking.[36] Empirically, this spontaneous symmetry breaking may be manifested by the alternating presence and absence of the tunneling conductance peak at $n = 0$ for two inequivalent sublattices in graphene, which has been previously demonstrated by STS studies of molecular graphene.[11] In this study, we also found that the point spectra of all strained regions exhibit statistically equal probabilities of the two zero modes. That is, the tunneling spectra at zero bias ($V = 0$) exhibit either a conductance peak or a conductance gap, as exemplified in Fig. S8A for the zero-bias conductance map of strained graphene over a Pd tetrahedron and the corresponding histograms in Fig. S8B. This finding therefore provides supporting evidence for spontaneous local time-reversal symmetry breaking due to strain-induced gauge potentials in real graphene.

*Nanoscale strain engineering of graphene-based valleytronic/spintronic devices*

The formation of periodic parallel graphene wrinkles by means of modern nanofabrication technology provides a pathway towards realizing controlled strain-induced effects for scalable development of graphene-based valleytronic devices. For instance, by patterning a valley-Hall device configuration with the long-axis parallel to graphene wrinkles as schematically illustrated in Fig. S9A, strong non-local resistance and valley-Hall effects may be detected under proper back-gated voltages, leading to a valley-Hall transistor similar to previous observation of the valley Hall effect in exfoliated monolayer graphene-on-*h*-BN flakes.[10] It is also conceivable to obtain highly valley-polarized currents through the combination of valley-splitters (Figs. 5A-5B) and valley-propagators (Fig. 5C), as conceptually illustrated in Fig. S9A. Furthermore, by injecting valley-polarized currents into strong spin-orbit coupled materials, the outgoing currents can become spin-polarized for spintronic applications (Fig. S9B). Finally, we note that many such devices can be developed by means of scalable and reproducible nanofabrication technology on large-area PECVD-grown graphene sheets,[32] thus making the applications of graphene-based nanoscale valleytronic/spintronic devices closer to reality.

*Strain-induced superconductivity in monolayer graphene*

The discovery of superconductivity in bilayer graphene twisted at a "magic angle"[37] has kindled great interest in exploring "flat-band" materials[38] (*i.e.*, materials with dispersionless energy-*vs.*-momentum relation) for induction of superconductivity. A recent theoretical proposal[39] suggests that superconductivity may be more easily realized in topological flat bands induced by strain in graphene through periodic ripples and by including the effect of electronic correlation. It is argued that the chiral *d*-wave superconductivity may be stabilized under strain even for slightly doped graphene, and that superconductivity thus derived could exhibit the long-sought-after superconducting states with non-vanishing center-of-mass momentum for Cooper pairs.[39]

In the limit of ($J/t$) ~ 1 where $J$ represents the antiferromagnetic coupling and $t$ is the nearest-neighbor hopping energy, the theoretical conditions necessary for the occurrence of superconductivity are found to be ($h/L$) ≥ 0.05 and $h^2/(La_0)$ ≥ 1, where $h$ and $L$ denote the height and periodic separation of the ripples, respectively, and $a_0$ ≈ 0.142 nm is the nearest carbon-carbon distance for equilibrium graphene.[39] For typical values of $h$ = 20 nm and $L$ = 300 nm in this work, we find ($h/L$) ≈ 0.067 and $h^2/(La_0)$ ≈ 9.39 so that both theoretical conditions are satisfied, implying possible occurrence of superconductivity if the premise of strong electronic correlation is justifiable.[39] Given this intriguing prospect, it would be worthwhile to empirically explore possible strain-induced superconductivity in monolayer graphene with architected parallel wrinkles and to verify the validity of strong electronic correlation under giant pseudo-magnetic fields. However, empirical verifications of superconductivity require measurements at cryogenic temperatures. Given that the thermal expansion coefficient for graphene is negative and those for typical substrate materials (such as silicon) are positive, the strain induced by architected substrates in graphene is expected to decrease with decreasing temperature. Therefore, proper consideration of such complications will be necessary in the investigation of possible strain-induced superconductivity in graphene.

In summary, we have demonstrated a controlled approach to manipulating the topological states in monolayer graphene via nanoscale strain engineering. By placing strain-free monolayer graphene on architected nanostructures to induce global inversion symmetry breaking, we are able to induce giant pseudo-magnetic fields (up to ~ 800 T) with desirable spatial distributions, realize global valley polarization, and achieve periodic one-dimensional topological channels for protected propagation of chiral Fermion modes in strained graphene. The methodology presented in this work not only provides a platform for designing and controlling the gauge potential and Berry curvatures in graphene, but is also promising for realizing scalable graphene-based valleytronic devices as well as strain-induced superconductivity.

## Materials and Methods

### Graphene/BN/Pd tetrahedron sample preparation

In this work, the Pd tetrahedron nanocrystals (NCs) were synthesized by a wet-chemical method.[33] The preparation procedure is briefly summarized below. We mixed 7.6 mg palladium (II) acetylacetonate (Pd(acac)$_2$), 16.5 mg iron (II) acetylacetonate (Fe(acac)$_2$), 50.0 mg polyvinylpyrrolidone (PVP), and 10.0 ml N, N-dimethylformamide (DMF) into a 30 mL vial. After ultrasonication for 5 minutes, the mixture was heated at 120 °C for 10 hours in an oil bath on a hotplate. The resulting precipitant were collected by centrifugation and rinsed with ethanol several times. A Si substrate was first ultrasonicated in acetone and subsequently in IPA for 10 minutes each, blown dry with dry nitrogen, and then loaded into a 100 W O$_2$-plasma for 5 minutes to remove any traces of hydrocarbon residue. The Pd tetrahedron suspension was dropped onto the Si substrate and spun at 1500 RPM for 1 minute. After spin-coating process, the sample was loaded into a 100 W O$_2$-plasma for 5 minutes again to remove any residue on the Pd tetrahedron NCs. The resulting typical size of the Pd tetrahedron NCs ranges from 50 to 70 nm and the height ranges from 40 to 60 nm, as exemplified in Figs. 1B and 1C. Next, a monolayer BN was transferred over the substrate covered by the Pd tetrahedron NCs, followed by the transfer of PECVD-graphene onto the BN/Pd-NCs. AFM and SEM measurements were performed on every step of the process. We found that granphene/BN conformed very well to the Pd tetrahedron NCs if they were well separated from each other, as exemplified by the AFM image in Figs. 1D and 1E (top panel). However, we found that graphene tended to form wrinkles along the Pd tetrahedrons if they were sufficiently closed to each other, as exemplified by the AFM images in the bottom panel of Fig. 1E. This situation is similar to our previous observation of graphene/$h$-BN on Au nanoparticles and graphene/$h$-BN on Si nanostructures.[31]

**Procedures for fabricating periodic arrays of graphene/h-BN/SiO$_2$ nano-cones**

SiO$_2$ nano-cones substrate fabrication is schematically shown in Fig. S4 and further described here. First, a typical electron beam lithography method was used to pattern an array of discs with ~ 50 nm diameters on a Si substrate with a 300 nm oxide layer. After development, 15 nm thick Ni is deposited and used as a mask in a $C_4F_8/O_2$ reactive ion etching (RIE) environment to create Si nano-pillars. After etching, the substrate was immersed in the buffered oxide etch for ~ 20 s until the Ni discs fell from the top of the nano-cones. A typical size of nano-cone is ~ 40 nm in diameter and ~ 20 nm in height. A monolayer $h$-BN was transferred over the SiO$_2$ nano-cones, followed by the transfer of PECVD-grown graphene.

**Scanning tunneling microscopic and spectroscopic studies of strain-engineered graphene**

Two types of monolayer strained graphene samples were investigated using STM/STS at room temperature. The samples were loaded onto our homemade STM system and pumped down to a vacuum level of $1.6 \times 10^{-6}$ torr. Atomically resolved topographic and spectroscopic measurements were carried out on samples at room temperature using a Pt/Ir STM tip with a typical tunnel junction resistance at 2 GΩ.

**Supplementary Materials**

Supplementary Notes (including Supplementary Figures S1 – S3)
Supplementary Figures S4 – S9

# References

1. Berry, M. V. Quantal phase factors accompanying adiabatic changes. *Proc. R. Soc. London, Ser.* A **392**, 45 (1984).
2. Mikitik, G. P. & Sharlai Yu. V. Manifestation of Berry's Phase in Metal Physics. *Phys. Rev. Lett.* **82**, 2147-2250 (1999).
3. Xiao, D., Chang M.-C. & Niu Q. Berry phase effects on electronic properties. *Rev. Mod. Phys.* **82**, 1959-2007 (2010).
4. Thouless, D. J., Kohmoto, M., Nightingale M. P. & den Nijs, M. Quantized Hall conductance in a two-dimensional periodic potential. *Phys. Rev. Lett.* **49**, 405 (1982).
5. Qi, X. L. & Zhang S. C. The quantum spin Hall effect and topological insulators. *Physics Today* **63**, 33-38 (2010).
6. Hasan, M. Z. & Kane C. L. Topological insulators. *Rev. Mod. Phys.* **82**, 3045-3067 (2010).
7. Castro Neto, A. H. *et al.* The electronic properties of graphene. *Rev. Mod. Phys.* **81**, 109-162 (2009).

8. Zhang, F., MacDonald A. H. & Mele E. J. Valley Chern numbers and boundary modes in gapped bilayer graphene. *Proc. Natl Acad. Sci.* USA **110**, 10546–10551 (2013).
9. Zhang, Y. *et al.* Experimental observation of the quantum Hall effect and Berry's phase in graphene. *Nature* **438**, 201–204 (2005).
10. Gorbachev, R. V. *et al.* Detecting topological currents in graphene superlattices. *Science* **346,** 448–451 (2014).
11. Gomes, K. K. *et al.* Designer Dirac fermions and topological phases in molecular graphene. *Nature* **483**, 306–310 (2012).
12. Ju, L. *et al.* Topological valley transport at bilayer graphene domain walls. *Nature* **520**, 650–655 (2015).
13. Xiao, D. *et al.* Coupled spin and valley physics in monolayers of $MoS_2$ and other group-VI dichalcogenides. *Phys. Rev. Lett.* **108,** 196802 (2012).
14. Mak, K. F., He K., Shan J. and Heinz T. F. Control of valley polarization in monolayer $MoS_2$ by optical helicity. *Nat. Nanotech.* **7,** 494 (2012).
15. Novoselov, K. S., Mishchenko A., Carvalho A., Neto A. H. C. 2D materials and van der Waals heterostructures. *Science* **353**, 6298 (2016).
16. Manzeli, S., Ovchinnikov D., Pasquier D., Yazyev O. V. and Kis A. 2D transition metal dichalcogenides. *Nat. Rev. Mat.* **2**, 17033 (2017).
17. Pearce, A. J., Mariani, E. and Burkard, G. Tight-binding approach to strain and curvature in monolayer transition-metal dichalcogenides. *Phys. Rev. B* **94**, 155416 (2016).
18. Li, H. *et al.* Optoelectronic crystal of artificial atoms in strain-textured molybdenum disulphide. *Nat. Comm.* **6**, 7381 (2015).
19. Reserbat-Plantey, A. *et al.* Strain superlattices and macroscale suspension of graphene induced by corrugated substrates. *Nano Lett.* **14**, 5044 (2014).
20. Tomori, H. *et al.* Introducing nonuniform strain to graphene using dielectric nanopillars. *Appl. Phys. Express* 4, 075102 (2011).
21. Choi, J. *et al.* Three-dimensional integration of graphene via swelling, shrinking, and adaptation. *Nano Lett.* **15**, 4525 (2015).
22. Pacakova, B. *et al.* Mastering the wrinkling of self-supported graphene. *Sci. Rep.* **7**, 10003 (2017).
23. Zhang, Y. *et al.* Strain modulation of graphene by nanoscale substrate curvatures: A molecular view. *Nano Lett.* **18**, 2098 (2018).
24. Jiang, Y. *et al.* Visualizing strain-induced pseudomagnetic fields in graphene through an h-BN magnifying glass. *Nano Lett.* **17**, 2839 (2017).
25. Mañes, J. L. Symmetry-based approach to electron-phonon interactions in graphene. *Phys. Rev. B* **76**, 045430 (2007).
26. Guinea, F., Katsnelson, M. I. and Vozmediano, M. A. H. Midgap states and charge inhomogeneities in corrugated graphene. *Phys. Rev. B* **77**, 075422 (2008).
27. Guinea F. *et al*. Energy gaps and a zero-field quantum Hall effect in graphene by strain engineering. *Nat. Phys.* **6**, 30–33 (2010).
28. Teague, M. L. *et al.* Evidence for strain-induced local conductance modulations in single-layer graphene on $SiO_2$. *Nano Lett.* **9**, 2542–2546 (2009).
29. Levy N. *et al*. Strain-induced pseudomagnetic fields greater than 300 Tesla in graphene nanobubbles. *Science* **329**, 544 (2010).
30. Yeh N.-C. *et al*. Strain-induced pseudomagnetic fields and charging effects on CVD-grown graphene. *Surf. Sci.* **605**, 1649–1656 (2011).
31. Yeh N.-C. *et al*. Nanoscale strain engineering of graphene and graphene-based devices. *Acta Mech. Sin.* **32**, 497 – 509 (2016).
32. Boyd D. A. *et al*. Single-step deposition of high-mobility graphene at reduced temperatures *Nature Communications* **6**, 6620 (2015).

33. Zhang Y. *et al*. Seedless growth of palladium nanocrystals with tunable structures: From tetrahedra to nanosheets. *Nano Lett.* **15**, 7519 (2015).
34. Lin W.-H. *et al*. A direct and polymer-free method for transferring graphene grown by chemical vapor deposition to any substrate. *ACS Nano* **8**, 1784 – 1791 (2014)
35. Lin W.-H. *et al*. Atomic-scale structural and chemical characterization of hexagonal boron nitride layers synthesized at the wafer-scale with monolayer thickness control. *Chemistry of Materials* **29**, 4700 − 4707 (2017).
36. Herbut, I. F. Pseudomagnetic catalysis of the time-reversal symmetry breaking in graphene. *Phys. Rev. B* **78**, 205433 (2008).
37. Cao, Y. *et al*. Unconventional superconductivity in magic-angle graphene superlattices. *Nature* **556**, 80 (2018).
38. Bistritzer, R. & MacDonald, A. H. Moiré bands in twisted double-layer graphene. *Proc. Natl Acad. Sci. USA* **108**, 12233–12237 (2011).
39. Xu, F. *et al*. Strain-induced superconducting pair density wave states in graphene. *Phys. Rev. B* **98**, 205103 (2018).
40. Neek-Amal M. and Peeters, F. M. Graphene on boron-nitride: Moiré pattern in the van der Waals energy. *Appl. Phys. Lett.* **104**, 041909 (2014).

**Acknowledgments**

The authors gratefully acknowledge joint support for this work by the Army Research Office under the MURI program (Award # W911NF-16-1-0472), National Science Foundation under the Physics Frontier Centers program for the Institute for Quantum Information and Matter (IQIM) at the California Institute of Technology (Award #1733907), and the Kavli Foundation.

**Author contributions**

N.-C. Yeh conceived the ideas and coordinated the research project. C.-C. Hsu synthesized and characterized the strain-free monolayer graphene, developed architected nanostructures, and transferred monolayer graphene and monolayer *h*-BN to the architected nanostructures for strain engineering. M. L. Teague performed the STM/STS studies on strained graphene and analyzed the topographic and spectroscopic data. J.-Q. Wang carried out the MD simulations to map out the strain-induced pseudo-magnetic fields and developed a semi-classical model to determine the trajectories of valley-polarized Dirac fermions. N.-C. Yeh wrote the paper with contributions from all coauthors.

**Competing interests**

The authors declare no competing interests.

**Data and materials availability**

All data needed to evaluate the conclusions in the paper are present in the paper and the Supplementary Materials. Additional data available from authors upon request.

# Supplementary Materials for
# Nanoscale Strain Engineering of Giant Pseudo-Magnetic Fields, Valley Polarization and Topological Channels in Graphene

C.-C. Hsu,[1]† M. L. Teague,[1]† J.-Q. Wang,[1]† N.-C. Yeh[1,2]*

[1]Department of Physics, California Institute of Technology, Pasadena, CA 91125, USA
[2]Kavli Nanoscience Institute, California Institute of Technology, Pasadena, CA 91125, USA

*Corresponding author: Nai-Chang Yeh (ncyeh@caltech.edu)
†These authors contributed equally to this work

**This file contains:**

Supplementary Notes (including Supplementary Figures S1 – S3)
Supplementary Figures S4 – S9

# Supplementary Notes

## Notes on molecular dynamics simulation

To estimate the pseudo-magnetic field induced by given underlying nanostructures in graphene, we employed the Molecular Dynamics (MD) method to compute the spatial distribution of the strain tensor. The numerical procedures are detailed below.

We began by creating a monolayer graphene sheet with a fixed number of carbon atoms and assuming that the positions of the boundary atoms remained invariant throughout the simulations. To induce distortions, we moved an underlying nanostructure adiabatically towards the graphene sheet until the desirable distortion was reached, and then relaxed the entire system until it reached equilibrium. The distorted positions of all carbon atoms were recorded and the spatially varying displacement field $\mathbf{u} \equiv u_x \hat{x} + u_y \hat{y} + h\hat{z}$ was extracted.

In the MD simulation, the interaction among carbon atoms was described by the AIREBO potential:[40]

$$E^{AIREBO} = \frac{1}{2} \sum_i \sum_{i \neq j} \left[ E_{ij}^{REBO} + E_{ij}^{LJ} + \sum_{k \neq i,j} \sum_{l \neq i,j,k} E_{ijkl}^{tors} \right] \quad (S1)$$

where $i, j, k$ and $l$ referred to individual atoms, $E^{REBO}$ is the part that explains the bonded interaction; $E_{ij}^{LJ}$ is Lennard-Jones potential that considers the non-bonded interaction, and $E^{tors}$ considers torsional interaction.

A layer of $h$-BN was inserted between graphene sheet and the underlying structures to minimize the perturbation from the substrate in experiment. This non-bonded effect between graphene and $h$-BN can be modeled by Lennard-Jones interaction:

$$E_{LJ} = 4\varepsilon \left[ \left( \frac{\sigma}{z} \right)^{12} - \left( \frac{\sigma}{z} \right)^{6} \right]. \quad (S2)$$

Here $z$ is the distance between the carbon atom and the $h$-BN layer (we neglect the in-plane energy variation since it is much smaller than the thermal energy at room temperature), the well depth $\varepsilon \approx 0.058 \, eV$ and the equilibrium distance $\sigma \approx 0.34 \, nm$.[40]

Assuming 2D elastic theory for continuum, the free energy induced by elastic lattice distortions may be expressed in terms of the elastic moduli and strain tensors by the following expression:

$$F_s = \int dx dy \left[ \frac{\kappa}{2} (\nabla_{2D} h)^2 + \frac{\lambda}{2} \left( \sum_i u_{ii} \right)^2 + \mu \sum_{i,j} (u_{ij})^2 \right], \quad (S3)$$

where $\kappa$ and $\lambda$ are the $c$-axis and in-plane bulk moduli of graphene, respectively, and $\mu$ is the in-plane shear modulus. Using Eq. (S3), the displacement field $(\mathbf{u}, \mathbf{r})$ may be determined by minimizing $F_s$.

Next, we consider the following scaling transformation:

$$\mathbf{r} \to \mathbf{r}' = k\mathbf{r}; \qquad h \to h' = kh(x,y), \tag{S4}$$

where $k$ is a dimensionless non-trivial constant and $h$ represents for the height variation of the underlying nanostructures. A new solution can be obtained for the system after scaling transformation by inserting Eq. (S4) into Eq. (S3) and then minimizing the total free energy. Noting that

$$\mathbf{u}(\mathbf{r}') = k\mathbf{u}(\mathbf{r}), \quad \Rightarrow \quad u'_{ij} = u_{ij}, \tag{S5}$$

we find that the strain tensor is invariant after the scaling transformation. Hence, the vector potential and pseudo-magnetic field of the scaled system become:

$$A'_x(\mathbf{r}') = \pm\frac{\beta}{a_0}(u_{xx} - u_{yy}) = A_x(\mathbf{r}), \quad A'_y(\mathbf{r}') = \mp\frac{\beta}{a_0}u_{xy} = A_y(\mathbf{r}); \tag{S6}$$

$$\mathbf{B}'(\mathbf{r}') = \nabla_{\mathbf{r}'} \times \mathbf{A}'(\mathbf{r}') = \nabla_{\mathbf{r}'} \times \mathbf{A}(\mathbf{r}) = \frac{1}{k}\nabla_{\mathbf{r}} \times \mathbf{A}(\mathbf{r}) = \frac{1}{k}\mathbf{B}(\mathbf{r}). \tag{S7}$$

Thus, the MD simulations may be carried out more effectively by numerical calculations in a smaller system with distortions of the same aspect ratios as those of a larger system, and then perform the scaling transformations in Eqs. (S4), (S6) and (S7) to achieve the results for an intended larger system.

Having outlined the general principles of MD simulations above, we explain in the following more details about our method and justification for using large scale topography to obtain the expected spatial distributions of the pseudo-magnetic fields, and then compare the computed results with experimental data.

We note that while theoretically the effect of strained-induced gauge potential given in Eq. (S6) may be directly deduced from the tensor components that are related to the displacement fields according to Eq. (6) in the manuscript, empirically it is not feasible to derive the strain-induced pseudo-magnetic field from the aforementioned expressions because the original position of each carbon atom is not traceable. Although in principle one may perform scanning tunneling spectroscopic studies to obtain the Landau levels at each spatial point and then compute the corresponding local pseudo-magnetic field, it is practically unrealistic to measure many millions of atomically resolved point spectra just to map out the spatial distribution of pseudo magnetic fields over an area of a few-hundred nanometers squared.

Fortunately, from MD simulations, we found that the pseudo-magnetic field largely comes from the out-of-plane lattice distortion. To illustrate this point, we demonstrate in the following series of figures our MD simulations for a monolayer graphene strained by two nanoparticles. Specifically, we note that the resulting in-plane lattice distortion ($du_x/dx$) is much smaller than the vertical lattice distortion ($dh/dx$). This observation implies that the out-of-plane component of the displacement field contributes the most to the resulting pseudo-magnetic field. Thus, by extracting the height variations from experimentally measured topographic data, we can obtain sufficiently accurate evaluations of the pseudo-magnetic fields over large areas.

As an example, we show below in Fig. S1 a plot of the topography for a monolayer graphene on top of two realistic nanoparticles from MD simulations:

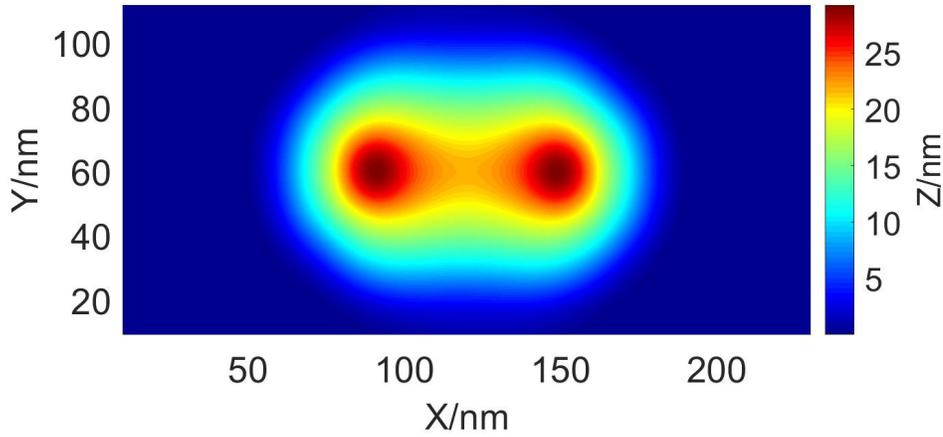

**Fig. S1. MD simulations of the topography of graphene on top of two nanoparticles.** The two nanoparticles considered for the simulations are based on realistic dimensions of Pd nanocrystals described in the main text.

From the above topographic data of $h(x, y)$, we can obtain the displacement fields, which yield both a map of $(\partial h/\partial x)$ and the map of $(\partial u_x/\partial x)$ shown in Fig. S2 below:

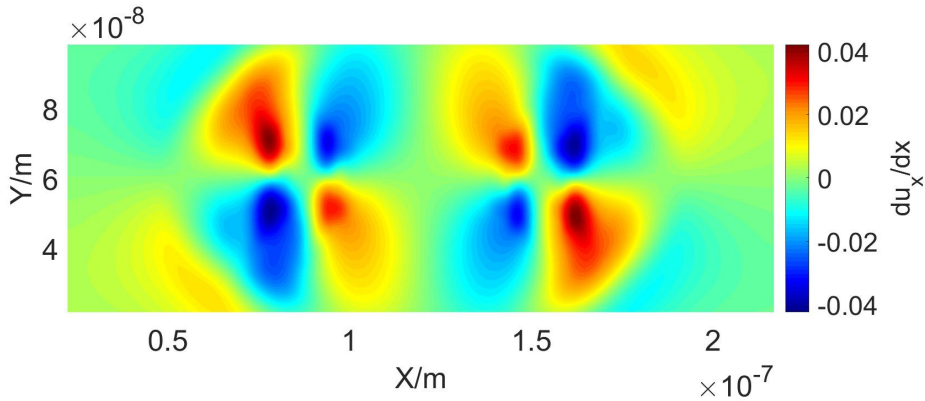

**Fig. S2. MD simulations of the strain map due to contributions from the in-plane distortion shown in Fig. S1.** Here the strain contributions associated with the in-plane distortion is given by $(\partial u_x/\partial x)$.

Next, we compare the contributions of the in-plane and out-of-plane distortions to the tensor component $u_{xx} \equiv (\partial u_x/\partial x) + [(\partial h/\partial x)^2/2]$ by plotting both the $[(\partial h/\partial x)^2/2]$ map and the $(\partial u_x/\partial x)$ map on the same color scale in Fig. S3A and Fig. S3B, respectively. It is evident that the contributions from $(\partial u_x/\partial x)$ to $u_{xx}$ in Fig. S3B is negligible when compared with those from $[(\partial h/\partial x)^2/2]$ in Fig. S3A.

166

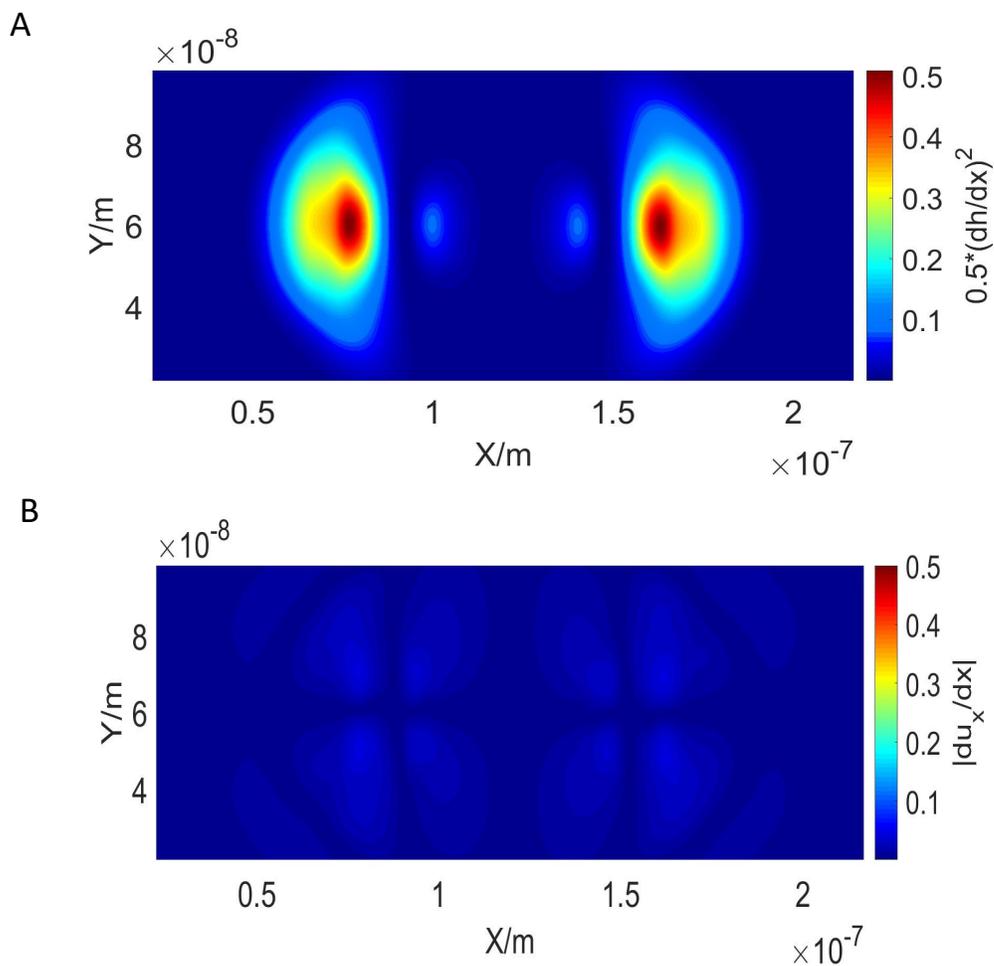

**Fig. S3. MD simulations of the strain maps due to the height and in-plane distortions.** (A) Strain map computed from the height distortion using the expression $[(\partial h/\partial x)^2/2]$. (B) Strain map computed from the in-plane distortion using the expression $(\partial u_x/\partial x)$. Here we note that the map of $(\partial u_x/\partial x)$ in (B) is the same as that in Fig. S2 except that the two maps are plotted using different scales.

# Notes on a semi-classical model for calculating the trajectory of valley-polarized Dirac fermions in the presence of strain-induced pseudo-magnetic fields

In the semi-classical model, the motion of electrons moving in a periodic lattice without external magnetic field is given by the following equations of motion:

$$\hbar \dot{\mathbf{k}} = e\mathbf{E}, \tag{S8}$$

$$\hbar \dot{\mathbf{r}} = \nabla_{\mathbf{k}} \varepsilon - \hbar \dot{\mathbf{k}} \times \mathbf{\Omega}(\mathbf{k}), \tag{S9}$$

where $\mathbf{E}$ is the applied electric field, $\varepsilon$ is the electron energy, and $\mathbf{\Omega}$ denotes the Berry curvature. For $K$-valley electrons moving in strained graphene, Eq. (S9) becomes:

$$\hbar \dot{\mathbf{r}} = \mathbf{v}_F - \hbar \mathbf{E} \times \mathbf{\Omega}(\mathcal{A}(\mathbf{r}), \mathbf{k}), \tag{S10}$$

where $v_F$ is the Fermi velocity, and the Berry curvature $\mathbf{\Omega}$ may be expressed as a function of the fictitious gauge potential $\mathcal{A}(\mathbf{r})$ induced by strain.

Alternatively, we may consider an electron moving in pseudo-magnetic field $\mathbf{B}_S(\mathbf{r})$ instead of computing the Berry curvature of valley electrons. Here $\mathbf{B}_S(\mathbf{r})$ can be obtained from the strain tensors using Eq. (2) in the manuscript and the relation $\mathbf{B}_S(\mathbf{r}) = \nabla \times \mathcal{A}(\mathbf{r})$. The semi-classical model that governs the transverse motion of electrons (relative to the direction of a uniform external electric field $\mathbf{E}$) may be described by the following equation of motion:

$$\hbar \dot{\mathbf{k}}_\perp = -e(\mathbf{v} \times \mathbf{B}_S(\mathbf{r})). \tag{S11}$$

Equation (S11) essentially attributes the rate of transverse momentum change for Dirac fermions in strained graphene to the effective spatially varying Lorentz force, where the magnitude of the velocity of Dirac fermions in graphene is comparable to the Fermi velocity of graphene:

$$v = v_F \approx 1 \times 10^6 \ m/s. \tag{S12}$$

Therefore, in the presence of a constant external electric field, Eqs. (S11) and (S12) may be rewritten into the following expression (after integrating over a finite time interval and neglecting the constant momentum shift due to the external electric field in the steady state):

$$d\mathbf{k}_\perp = -\frac{eB_S}{\hbar}(d\mathbf{R} \times \hat{z}), \tag{S13}$$

where $\hat{z}$ is the unit vector along the pseudo-magnetic field direction, and $\mathbf{R}$ is the position of the $K$-valley Dirac fermion in real space. Since the cross product of $\hat{z}$ and $\mathbf{R}$ basically rotates $\mathbf{R}$ by 90° inside the plane of motion, the trajectory of electron in real space is simply (or part of) its orbit in $\mathbf{k}$-space scaled by the factor $(eB_S/\hbar)$. The time-average magnitude of $\mathbf{k}$ remains constant because $\mathbf{B}_S$ is perpendicular to the plane of motion. Therefore, we obtain the radius of curvature $\rho(t)$ (for the $K$-valley electron trajectory as follows:

$$\rho(t) = -\frac{\hbar}{eB_S(x,y)}|K|. \tag{S14}$$

To evaluate the validity of the aforementioned semi-classical model, let's make a quick estimate of the scale of electron motion in strained graphene using Eq. (S14). For valley electrons in graphene, we have $|K|=|K'|=\frac{4\pi}{3\sqrt{3}a}$. Thus, for a typical pseudo-magnetic field $B_S = 300\,T$ in our experiments, we find $\rho$ = 40 nm from Eq. (S14), which is much larger than the corresponding magnetic length $l_m = \sqrt{\hbar/eB_S} \sim 1.4\,nm$. Therefore, the semi-classical description of trajectory is justifiable for Dirac fermions in strained graphene.

Given Eq. (S14), we may define the cyclotron effective mass for the motion of Dirac fermions in graphene under strain-induced pseudo magnetic fields as follows:

$$m_c^* = \frac{eB_S\rho}{v_F}. \tag{S15}$$

We note that the cyclotron mass defined in Eq. (S15) is different from the effective mass obtained from band structure, for valley electrons under 300T magnetic field, $m_c^*$ is about twice as big as the free electron mass, implying significant localization of Dirac fermions in the presence of large pseudo magnetic fields.

Now that we have justified the semi-classical description and obtained an effective mass for the Dirac fermions in strained graphene, we may express the equation of motion for the Dirac fermions by the following relation:

$$e\dot{\mathbf{r}} \times \mathbf{B}_S(\mathbf{r}) = m_c^* \ddot{\mathbf{r}}. \tag{S16}$$

Thus, the trajectory of electron motion $\mathbf{r} = \mathbf{r}\,(t)$ can be derived from solving Eq. (S16).

# Supplementary Figures S4 – S9

## Supplementary Figure S4

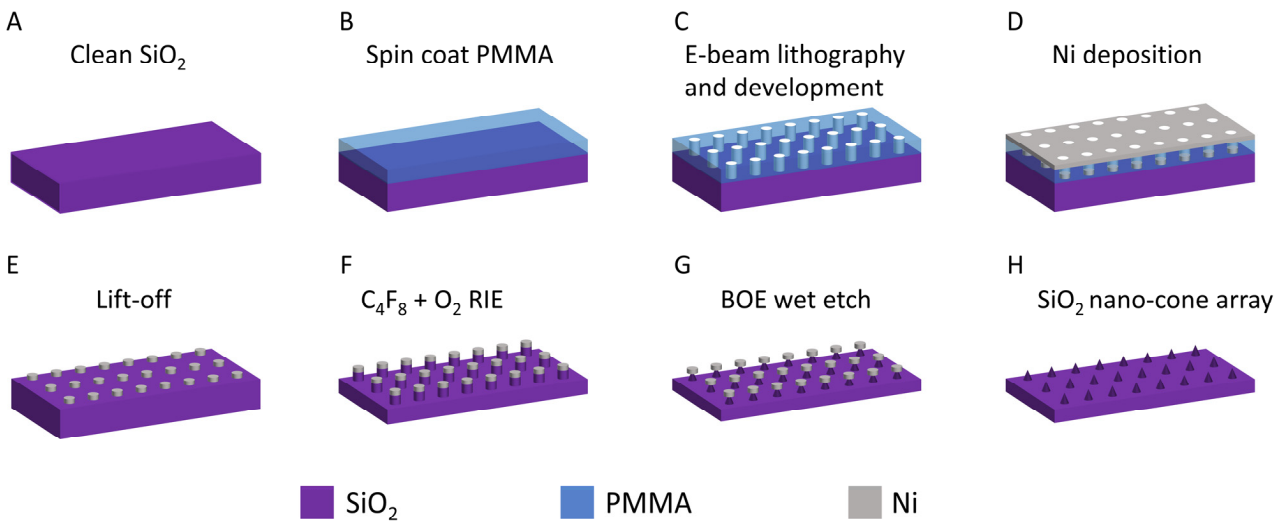

**Fig. S4. SiO₂ nano-cone array fabrication process.** (A) Si chip with a 300 nm oxide layer was ultra-sonicated in acetone and IPA for 10 min respectively, and then blown dry with nitrogen. (B) Spin coat ~ 100 nm PMMA on the SiO$_2$ and bake on a hot plate at 180 °C for 1 minute. (C) E-beam lithography and development. (D) 15 nm Ni deposition. (E) Lift off the resist by soaking the chip in acetone overnight. (F) Use C$_4$F$_8$/O$_2$ reactive ion etching (RIE) to create SiO$_2$ nano-pillars. (G) Dip the chip in buffered oxide etch (BOE) for ~ 20 seconds until Ni discs fall off. (H) SiO$_2$ nano-cone array.

## Supplementary Figure S5

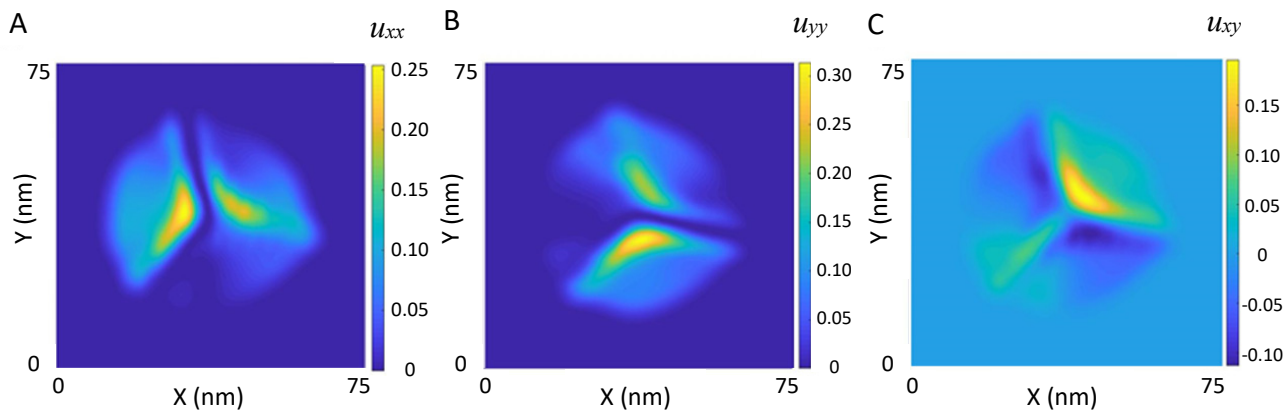

**Fig. S5. Maps of strain tensors induced by an underlying Pd nanocrystal on a monolayer graphene sheet:** (A) $u_{xx}$; (B) $u_{yy}$; and (C) $u_{xy}$.

## Supplementary Figure S6

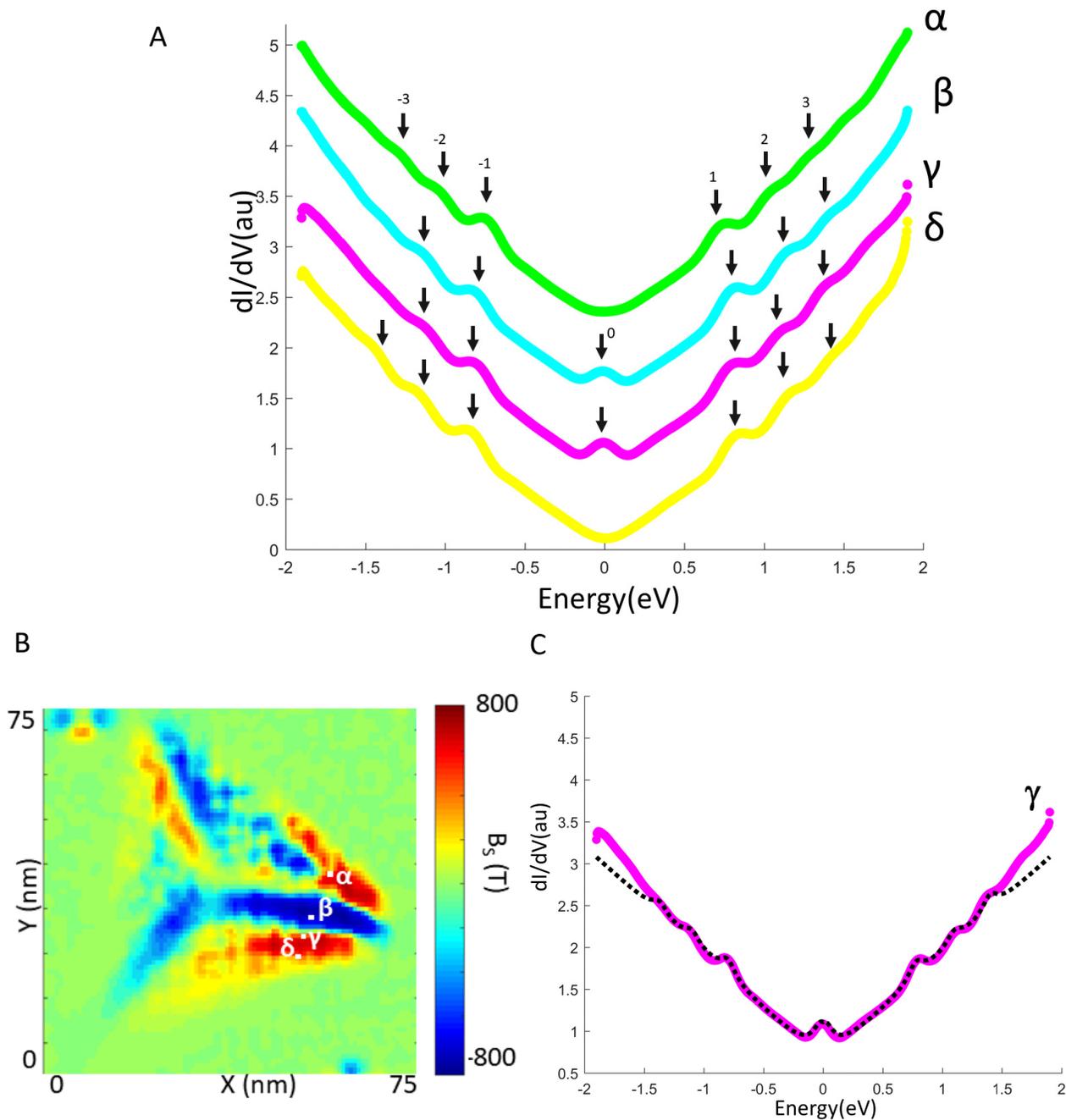

**Fig. S6. Strained-induced pseudo-magnetic fields and the resulting spectral Landau levels in strained graphene.** (A)-(B) Four representative point spectra ($\delta, \gamma, \beta, \alpha$) in (A) taken on strained graphene locations indicated on the pseudo-magnetic field map in (B), showing strain-induced spectral Landau levels with $n = 0, \pm1, \pm2$ and $\pm3$. Here we note that two of the spectra exhibit the zero-mode Landau level whereas the other two show absence of the zero mode, which are the consequences of local time-reversal symmetry breaking. (C) Theoretical fitting (black dashed curve) to the point spectrum $\gamma$ with a pseudo-magnetic field value $|B_S| = 592$ T, showing good agreement up to $n = \pm3$. The theoretical fitting curve is obtained as follows: First, following the analysis described in Ref. 28, we fit the point spectrum to a background Dirac curve that also incorporates a small energy gap (~ 23 meV) associated with the contribution from an out-of-the-plane phonon mode at the zero bias. Next, all strain-induced spectral peaks that deviate from the

Page 9 of 11

Dirac curve are identified and analyzed using Eq. (3) in the manuscript to obtain the corresponding Landau level indices $n$ with a single pseudo-magnetic field value $|B_S|$ among all Landau levels. Each Landau level after the subtraction of the Dirac curve background is fit to a Lorentzian curve, and then all Lorentzian curves for the Landau levels with a single value of the pseudo-magnetic field $|B_S|$ combined with the background Dirac spectrum form the theoretical fitting shown as the black dashed curve.

**Supplementary Figure S7**

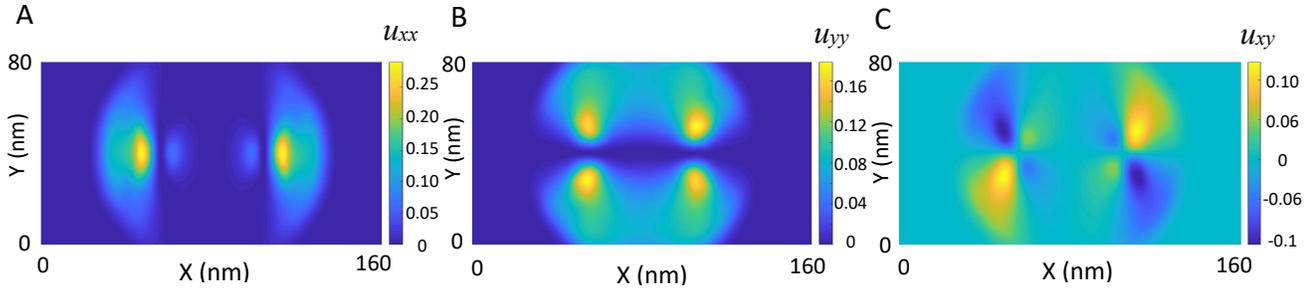

**Fig. S7. Maps of strain tensors induced by two underlying Pd nanocrystals on a monolayer graphene sheet:** (A) $u_{xx}$; (B) $u_{yy}$; and (C) $u_{xy}$.

**Supplementary Figure S8**

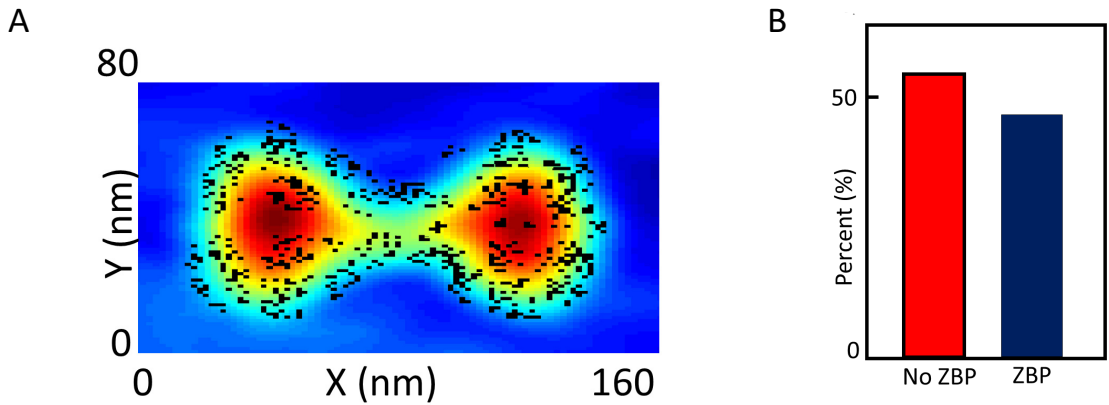

**Fig. S8. Scanning tunneling spectroscopic evidence for strain-induced spontaneous local time-reversal symmetry breaking and two zero modes in monolayer graphene.** (A) Spatially resolved map of the locations showing finite zero-bias conductance peaks (marked in black squares) in monolayer graphene strained by two Pd-tetrahedron NCs shown in Fig. 3B. (B) Histogram of the occurrence of zero-bias conductance peaks (denoted as "ZBP") and gaps (denoted as "No ZBP") for spectra taken in monolayer graphene strained by two Pd-tetrahedron NCs shown in Fig. 3B, showing statistically comparable probabilities for the appearance and absence of the zero-bias conductance peaks in the strained monolayer graphene.

## Supplementary Figure S9

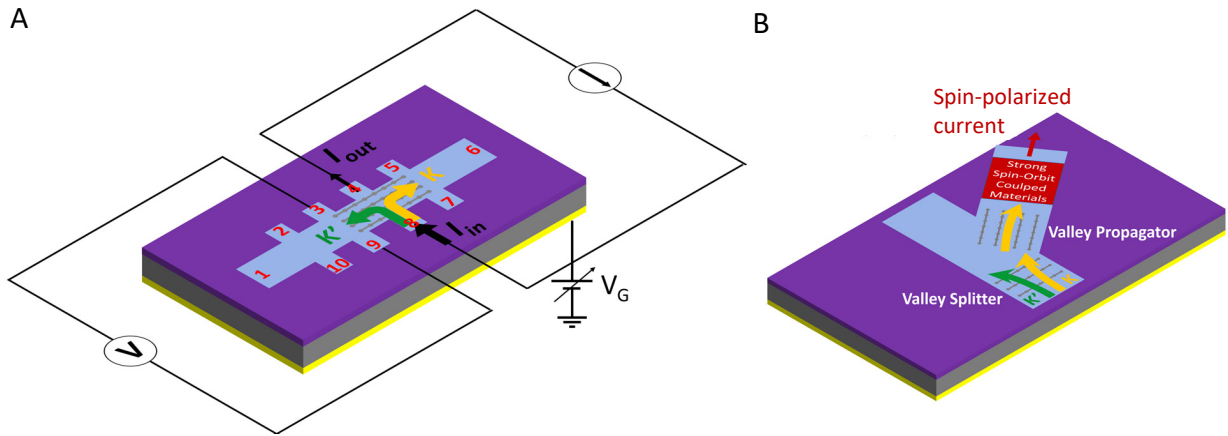

**Fig. S9. Schematics of the experimental configurations for graphene-based valleytronics and spintronics.** The blue-shaded area represents a patterned monolayer graphene/$h$-BN sample with strain-engineered periodic parallel wrinkles, and the purple region denotes a thin-layer of $SiO_2$ on top of a Si-substrate (grey region). (**A**) A graphene valley-Hall transistor:[10] For graphene wrinkles parallel to the long axis, an incident current ($I_{in}$) perpendicular to the wrinkles will lead to splitting of the K and K′ Dirac fermions. Therefore, in addition to the longitudinal resistance ($R$) that may be determined either from $R = (V_{25}/I_{16})$ or equivalently from $(V_{10,7}/I_{16})$, a non-local resistance ($R_{NL}$) may be detected from $R_{NL} = (V_{39}/I_{48})$ as shown above, or equivalently from $(V_{57}/I_{48})$. By placing the graphene Hall bar on the $SiO_2$/Si substrate and attaching a back gate to the Si, the Fermi level of the graphene can be controlled relative to the Dirac point by tuning the gate voltage ($V_G$) so that a sharp peak in $R_{NL}$-vs.-$V_G$ is expected when the Fermi level coincides with the Dirac point.[10] This configuration is therefore a field effect transistor. (**B**) A valleytronic-to-spintronic device: The combination of graphene-based valley-splitters and valley-propagators can lead to the generation of valley-polarized currents, as schematically illustrated by the yellow arrows for the trajectory of K-valley Dirac fermions. The injection of valley-polarized currents into a strong spin-orbit-coupled material can further lead to outgoing spin-polarized currents for spintronic applications.